\documentclass{amsart}
\usepackage{amssymb}
\usepackage{graphicx}
\usepackage{epstopdf}
\newcommand{\CF}{\hbox{{$\mathcal F$}}}

\newcommand{\R}{\mathbb{R}}

\newcommand{\C}{\mathbb{C}}

\newcommand{\note}[1]{}
\newcommand{\h}{{\scriptstyle\frac{1}{2}}}
\newcommand{\extd}{{\rm d}}

\newcommand{\eps}{{\epsilon}}
\newcommand{\tens}{\mathop{\otimes}}

\newcommand{\<}{\langle}
\renewcommand{\>}{\rangle}
\newcommand{\del}{\partial}

\newcommand{\trace}{{\rm Tr}}

\newcommand{\lcross}{{>\!\!\!\triangleleft}}

\newcommand{\rlbicross}{{\triangleright\!\!\!\blacktriangleleft}}
\newcommand{\lrbicross}{{\blacktriangleright\!\!\!\triangleleft}}

\newcommand{\eqn}[2]{\begin{equation}#2\label{#1}\end{equation}}

\begin{document}

\title[Spontaneous time generation and Planckian bound]{Noncommutative model with spontaneous time generation and Planckian bound}
\author{S. Majid}
\address{School of Mathematical Sciences\\
Queen Mary, University of London\\ 327 Mile End Rd,  London E1
4NS, UK\\ \& 
Perimeter Institute for Theoretical Physics\\
31 Caroline St N., Waterloo, ON N2L 2Y5, Canada}



\maketitle

\begin{abstract}
We illustrate the thesis that if time did not exist, we would have to create  it if space is noncommutative, and extend functions by something like Schroedinger's equation. We propose that the phenomenon is  a somewhat general mechanism within noncommutative geometry for  `spontaneous time generation'. We  show in detail how this works for the $su_2$ algebra $[x_i,x_j]=2\imath\lambda \eps_{ij}{}^kx_k$ as noncommutative space, by explicitly adjoining the forced time variable. We find the natural  induced noncommutative Schroedinger's equation and show that it has the correct classical limit for a particle of some mass $m\ne 0$, which is generated as a second free parameter by the theory.  We show that plane waves exist provided $|\vec p|< \pi/2\lambda$, i.e. we find a Planckian bound on spatial momentum. We also propose dispersion relations $|{\del p^0\over\del \vec p}|=|\tan({\lambda}|\vec p|)|/m\lambda$ for the model and explore some elements of the noncommutative geometry. The model is complementary to our previous bicrossproduct one.  \end{abstract}

\section{Introduction}

The origin of a time direction is a fundamental issue in any theory of quantum gravity, as likewise is the origin of mass for elementary particles. In this article we point out that previously known results on noncommutative differential calculi on quantum algebras can be viewed as evidence for a general phenomenon which we call `spontaneous time generation' in which both time and non-zero mass can be created  by  even a small amount of  noncommutativity in  space  (or more precisely in its geometry). Put another way, a noncommutative deformation of space by a parameter $\lambda$ can induce its own canonical evolution, forced by nothing other than the most minimal assumptions on existence of a differential structure on the space.

It is important that we use here the absolutely minimal and generally accepted notion of differential calculus or `exterior algebra' applicable to a noncommutative algebra $A$ and common to all main approaches, i.e.  we will not put in anything beyond this `by hand'. This is to specify a bimodule of `1-forms' $\Omega^1$ with left and right multiplication by a `function' in $A$, and a $\extd: A\to \Omega^1$ operation obeying the Leibniz rule
\[ \extd (ab)=a\extd b+ (\extd a)b.\]
One also requires that elements of the form $a\extd b$ span $\Omega^1$ and (a connectedness condition) that $\extd a=0$ if and only if $a$ is a constant. These are minimal properties that nevertheless suffice to do basic gauge theory on any algebra.

Next we note that for noncommutative algebras there is typically an element $\theta\in \Omega^1$ that generates the calculus in the sense
\[     [\theta, a]=\lambda  \extd a\]
where $\lambda$ is a parameter controlling the noncommutativity of the algebra $A$. Both sides go to zero when $\lambda\to 0$ (because classical 1-forms commute with functions), so this is a purely `quantum' phenomenon. The reason for the quotation marks here is that in our application $\lambda$ is not related to $\hbar$ but is a new parameter in physics controlling a possible noncommutativity in space or spacetime, or equivalently `cogravity' as curvature in momentum space\cite{Ma:mea}.   The element $\theta$ must exist for example   in the case of a semisimple Hopf algebra and a translation-covariant calculus, but also more generally whenever the geometry of $A$ is fully noncomutative. On the other hand there is no reason at all to have such an element in the classical case as the equation above is empty.  Therefore as we deform the classical algebra and its geometry, the constraints of noncommutative geometry will force the existence of a generating 1-form $\theta$ which {\em will have no reason to have any classical counterpart}. We are going to interpret this element $\theta$ as a time direction and the phenomenon then is that there may or may not be such a direction $\extd t$ in our classicial geometry with  $\theta$ its deformation, but if there is not such a time direction it will have to be adjoined. This is what we propose here as   `spontaneous generation' of time in our quantum groups approach to noncommutative geometry. Moreover, the partial derivative $\del^0$ associated to this $\theta=\extd t$ will generally have a classical limit for $\del^0/\lambda$ and this will be our  induced {\em classical} Hamiltonian generated by the noncommutative geometry.  Finally, we are free to change the normalisation of $\theta$ in the above, which appears then as a new parameter induced at the same time.

In this article we will demonstrate this phenomenon in detail for one of the most basic noncommutative 3-dimensional spaces, namely the angular momentum space 
\eqn{fuzzyR3}{[x_i,x_j]=2\imath\lambda\eps_{ij}{}^k x_k}
 of which the noncommutative geometry was studied in \cite{BatMa:non}. This is the usual quantisaiton of $\R^3$ as the coadjoint space $su_2^*$ with its Kirillov-Kostant Poisson bracket, denoted $\R^3_\lambda$. The need  for an extra direction $\theta$ in the calculus, as well as the link with Schroedinger's equation was proposed here, and in  this sequel we develop it in detail. Section~2 is a reprise of the noncommutative differential geometry on (\ref{fuzzyR3}) that we need from \cite{BatMa:non} except that we leave the normalisation $\theta$ as a free nonzero parameter $\mu$.  Section~3 gives a quick account of quantum group Fourier transform and a direct derivation of formulae for the partial derivatives on 3D plane waves, which will also be needed. 
 
 In Section~4 we proceed to explicitly adjoin a new variable $t$, commuting with $x_i$, such that $\theta=\extd t$. In this extended space-time algebra the natural equation
\eqn{ncse}{ \del^t \psi(x,t)=0}
(or that $\extd \psi$ is purely spatial) is now our proposed noncommuative Schroedinger's equation (NCSE). We show that in the limit $\lambda\to 0$ this reproduces  the usual  Schroedinger equation for $\psi$ for a particle of mass $m$ with Compton wavelength $\mu=1/m$. {\em We work throughout in units $\hbar=c=1$}. In fact it turns out that as $m$ approaches $1/\lambda$  we have to the next order a noncommutave version of the 4D  Euclidean wave equation for $e^{-\imath m t}\psi(x,t)$, suggesting a Euclidean aspect to the theory. Our construction is, however, very much tied to the nonrelativistic coordinate system consisting of $x_i$ space and the induced $t$ although it does not preclude the possibility of a Lorentz or 4D Euclidean  (quantum) group action in the deformed theory. There is at least  a full spatial Euclidean group of motions preserved in the construction, as the quantum group double $D(U(su_2))$ \cite{BatMa:non}. In Section~5 we finally solve the NCSE on plane waves, with solutions $e^{\imath \vec p\cdot x}e^{\imath p^0t}$ of energy
 \eqn{energy}{ p^0=-{1\over m\lambda^2}\ln(\cos({\lambda}|\vec p|))}
 {\em provided} $|\vec p|< \pi/2\lambda$.  
 We also find the group velocity computed naively by differentiation of (\ref{energy}). It is a deformation of $|\vec p|/m$. 
 
Our model comes out of noncommutative geometry and is not tied to specific values of $\lambda,m$. Our own view is that the theory could also be applicable in certain circumstances as an effective description of position `fuzziness' within quantum mechanics at the Compton wavelength scale of an ordinary particle, in the spirit of \cite{Pen}. However, if the theory appeared as a next-to-classical effective theory in a theory of quantum gravity, $\lambda$ would be expected to be of the order of the Planck scale and hence we would have a Planckian spatial momentum cut-off for our particle. 
In fact it is known that the quantum double $D(U(su_2))$ controls the tensor products of certain states in 2+1 quantum gravity \cite{BaiMul,Sch:com}, therefore the result  \cite{BatMa:non}  that this quantum group acts covariantly  on  $\R^3_\lambda$   suggested that the latter should indeed be the effective noncommutative space in the next-to-classical approximation of 2+1 Chern-Simons quantum gravity. The algebra $\R^3_\lambda$  has also  been proposed in  string theory and in the reduction of certain matrix models but more usually in this context projected to a matrix algebra by setting the casmir equal to $j(j+1)$, which is to say a `fuzzy sphere'\cite{ARS, BMC}. However, we do not make such a projection here and we do not use the ad-hoc derivations-based matrix methods previously used for such objects.  The noncommutative geometry in our case and with time adjoined is actually very rich and explored in Section~6 where we show the existence of a closed radial polar coordinate system and some elements of gauge theory. 
 
 We note also that a Planckian-cutoff is already a feature of the bicrossproduct spacetime \cite{MaRue:bic} and we provide a comparison with this older model in the Appendix. In fact it has been known for some time that the zero-mass shell equation in the bicrossproduct model is deformed to 
  \[ |\vec p|={1\over\lambda}(1-e^{-\lambda p^0}),\quad {\rm or}\quad p^0=-{1\over\lambda}\ln(1-\lambda |\vec p|),\]
see (\ref{kappacas}) in the appendix, which (obviously) has the bound $|\vec p|< 1/\lambda$.   Recently, some authors have dubbed models with such a Planckian cut-off as `doubly special' under the claim that the asymptotic feature of special relativity is now `doubled' by this additional asymptotic bound \cite{KowNow}. While debatable, our own view is that such rebranding of the bicrossproduct model  (which remains the main model in the theory)  is unjustified as we also explain in the appendix:  what actually happens in our view is that the usual mass-shell hyperboloid is deformed nonlinearly and what used to be a 45-degree cone is now bent into a cylinder with vertical walls, rather than being a new bound.  We also outline a different point of view on the cut-off in line with \cite{Ma:pla} (where $p$ was viewed as position space rather than momentum) as an event-horizon-like coordinate singularity.  In the present model at least part of the reason for a cut-off is rather more transparent: momentum space is compactified to a sphere $S^3=SU_2$ according to the quantum group Fourier transform.

Returning to the general phenomenon of spontaneous time creation, we note that for $\C_q[SU_2]$  it is again known that the smallest bicovariant calculus is 4D not 3D and that the extra direction $\theta$ is linked to the Laplace operator, now on the quantum group as a noncommutative $S^3$. It is also already known that the local 4D cotangent bundle indeed has a natural $q$-Minkowski space metric, see \cite{Ma:book} for a review and, for example \cite{GomMa:non} for full calculations at $q$ a root of unity (which is likely to be the physical case if such a deformation arises from quantum gravity in view of the well-known role of this quantum group in CFT). The consistent addition of a time variable, however,  and an analysis along the lines of the present paper is somewhat technical and will be presented elsewhere. 

Finally, let us note that our proposal on time has no relation that we are aware of to the modular group in the theory of von-Neuman algebras; our results are purely algebraic and not connected with functional analysis. Briefly, Tomita and Takesaki in the 1970s showed that every von-Neumann algebra carries with it a 1-parameter automorphism group $\sigma_t$ generated by the positive part $\Delta$ in the polar decomposition of the $*$-operation relative to a state. Translation by a finite imaginary interval in $t$ is used to characterise KMS states in equilibrium quantum statistical mechanics on the algebra. Hence some authors, notably\cite{ConRov}, have proposed that $t$ here should be viewed as time canonically associated to the von-Neumann algebra in a suitable setting. While  such a point of view is interesting, the phenomenon  we propose in the `quantum groups approach' to noncommutative geometry is a rather more concrete one  in which we shall show that a parameter $t$ has to be adjoined and wave functions with respect to it naturally obey Schroedinger's equation for some mass $m$. Our theory at the present level does not determine $m$, only that it must be non-zero, although $m=1/\lambda$ does present some simplifications in the mathematical structure as noted above. 

According to our analysis this mass and time generation is forced by the axioms of a differential calculus. We mention one alternative, which is to change these axioms by giving
up associativitiy.  At the semiclassical level it corresponds to curvature of an underlying Poisson-compatible preconnection\cite{BegMa:twi}. (If one simply drops the $\theta$ terms in the calculus one would have such a situation with Poisson-curvature and nonassociativity appearing at order $\lambda^{2}$). This gives an alternative idea of the nature of the obstruction involved. In physical terms one could say that the spatial translation group (as expressed in the differentials) is `anomalous' under the process of deformation quantisation, with anomaly controlled by the above Poisson-curvature and nullified by adding an extra dimension.  

\section{Reprise of differential calculus and plane waves on $\R^3_\lambda$}

Since the calculus is crucial to our entire analysis, let us briefly reprise the construction in \cite{BatMa:non} . For Lie groups it is well-known that the translation-invariant differential structure is unique; for quantum groups there is a parallel theory for the weaker minimal axioms above that translation-invariant $\Omega^1$ are freely generated over their space of invariant 1-forms and this latter space can be
classified in terms of ideals in the augmentation ideal (the kernel of the counit) of the Hopf algebra. 
See \cite{Ma:dcalc} for a modern review of the theory. In general we refer to \cite{Ma:book} for the notations and more basic theory of Hopf algebras.

Of course, $\R^3_\lambda$ is an additive Hopf algebra (in fact a classical enveloping algebra) with
\[ \Delta x_i=x_i\tens 1+1\tens x_i,\quad \eps x_i=0,\quad Sx_i=-x_i,\]
so we may use Hopf algebra or quantum group methods, and we do.  The augmentation ideal in our case is the subalgebra  $U(su_2)^+$ generated  by the $x_i$ but not including $1$. So left-invariant calculi (which will automatically be bicovariant in the present context) will be classified by ideals in here. These in turn are given by the kernels of matrix representations, more precisely by pairs $(\rho,v)$ consisting of a representation and a ray in the representation space. The kernel of the map consisting of applying $\rho$ to $v$ is the ideal we need, and the  left-invariant forms become identified with the orbit of $v$ (which is the whole representation space for a cyclic vector). In our case, one can compute the differential calculi for the spin 0, 1/2 and 1 representations, of dimension 1,2,3 respectively. Especially, the last of these might be expected to be the correct calculus but in all these cases one may compute that $\extd $ has a large kernel so these calculi are not successful. The next smallest is ${1\over 2}\tens {1\over 2}$ which is to say the Pauli matrix representation on $M_2(\C)$ where the algebra acts from the left and from the right (a 4-dimensional representation) by Pauli-matrices. The canonical vector $v$ is the identity matrix. This 4D calculus as shown to fulfill the connectedness property and is as we see the smallest such. We refer to \cite{BatMa:non} for details. We are assuming that $\lambda\ne 0$ (otherwise we could have the classical calculus). 

 The resulting calculus has commutation relations
\[\extd x_i=\lambda \sigma_i,\quad   x_i \theta - \theta x_i = \imath{\lambda^2\over\mu}
\extd x_i,\]
\[ (\extd x_i)
x_j-x_j \extd x_i=\imath\lambda\eps_{ij}{}^k\extd
x_k+\imath\mu\delta_{ij}\theta, \] where $\theta$ is a multiple of the $2\times 2$ identity matrix and,
together with the Pauli matrices $\sigma_i$, completes the basis of left-invariant 1-forms.  In the present paper relative to \cite{BatMa:non} we have put in a critical factor of $\imath$ and scale factor $\mu\ne 0$ in the  normalisation of $\theta$  to express explicitly that we are free to chose this normalisation. One might expect $\mu\sim \lambda$ if both are generated by some deeper theory, or one might consider $\mu$ as second and independent length scale in the theory. The factor $\imath$ is justified as follows: to speak about unitarity all our algebras will be $*$-algebras and thinking of the $x_i$ as observables in the quantum algebra and real functions in the classical limit, we require
\[ x_i^*=x_i,\]
which is consistent with $\lambda$ real for the conventions used for $\R^3_\lambda$. Next, it is reasonable that $(\extd x_i) ^*=\extd x_i$ if we want these also to be observables and to be identified with real 1-forms in the classical limit. The entire exterior algebra is generated by 1-forms, our case with the usual anticommutative wedge product, and we require this to be a $*$-algebra with
\[ (\extd \alpha)^*=(-1)^{|\alpha|}\extd (\alpha^*)\]
for a form of degree $|\alpha|$. These conventions are not always adhered to in the literature (there are other equivalent ones in other contexts) but at least here where there is a clear match with the classical limit they are reasonable. For example, one of the main things one does with differential 1-forms is gauge theory. If $u$ is unitary then 
\[ (u^{-1}\extd u)^*=(\extd u)^*u=(\extd u^{-1})u=\extd(1)-u^{-1}\extd u=-u^{-1}\extd u\]
is antihermitian. So connection 1-forms $\alpha$ are antihermitian. Then the curvature obeys
\[ F(\alpha)^*=(\extd\alpha+\alpha\wedge\alpha)^*=-\extd\alpha^*+\alpha^*\wedge\alpha^*=F(\alpha)\]
which is to say behaves homogeneously under $*$. This justifies our reality conventions for the calculus. Indeed, a $U(1)$ connection in such a (trivial bundle) gauge theory would be $\imath$ times a real 1-form in classical geometry. 

For our purposes we likewise require that $\theta^*=\theta$ i.e. a real 1-form in the classical limit, which determines the normalisation used. An alternative is to require that $\theta^*=-\theta$ which would be more suitable for applications in which $\theta$ has the interpretation of a reference connection. The latter was the convention and application for $\theta$ in \cite{BatMa:non} for example.

Once one has fixed the differential calculus,  the partial derivatives $\del^i$ as operators on $\R^3_\lambda$ are completely determined by \[ \extd \psi(x)=(\del^i \psi)\extd
x_i+(\del^0 \psi)\theta\]  
They are not derivations (that would be an older and widely discredited approach to noncommutative geometry); rather they are braided-derivations with respect to a certain solution of the Yang-Baxter
equations (induced from the quantum double). 

Finally plane waves take the form of group elements in the enveloping algebra $\R^3_\lambda$,  
\[ \psi_{\vec p}(x)=e^{\imath  \vec p\cdot x},\quad \vec p\in \R^3\]
The momenta $p_i$ are nothing but local coordinates for the corresponding point $e^{\imath{\lambda}\vec p\cdot\sigma}\in SU_2$ where ${\lambda}\sigma$ is the representation by Pauli matrices. It is really elements of this curved space $SU_2$ where momenta live, as evident in the addition law for momenta determined by the plane waves:
\[ \psi_{\vec p}.\psi_{\vec p'}=\psi_{C(\vec p,\vec p')}\]
where $C(\vec p,\vec p')$ is the Campbell-Baker-Hausdorf series. This is the general procedure for any enveloping algebra of a Lie algebra and the one we use here. Other coordinate systems are also possible, for example by Euler angles. We will show next that  our plane waves are eigenfunctions of the $\del^i$. The result is in \cite{BatMa:non} but the proof now is entirely different and self-contained.

\section{Quantum group Fourier transform and action on plane waves}

The algebra $\R_\lambda^3=U(su_2)$ has dual $\C[SU_2]$ and Hopf algebra Fourier transform (after suitable completion) takes one between these spaces. Thus, in one direction   
\[ \CF(f)=\int_{SU_2}\kern-10pt \extd u f(u) u\approx \int_{\R^3} \kern -5pt \extd^3p\, J({\vec p})f(\vec p)e^{\imath \vec p\cdot x}\]
for $f$ a function on $SU_2$. We
use the Haar measure on $SU_2$. The local result on the right has $J$ the Jacobian for the change to the local $\vec p$ coordinates and $f$ is written in terms of these.  Differential operators on $\R_\lambda^3$ are given by the action of elements of $\C[SU_2]$ and are diagonal on these plane waves,
\[ f. \psi_{\vec p}=f(\vec p)\psi_{\vec p},\]
which corresponds under Fourier transform  simply to pointwise multiplicaiton in $\C[SU_2]$.  
This quantum group Fourier transform approach to noncommutative geometry whereby it becomes
equivalent to a theory of classical but {\em curved} momentum space was introduced by the author in \cite{Ma:ista} more than a decade ago. We refer to \cite{Ma:mea} for a more recent review. Of course Fourier transform by other more conventional methods such as spherical harmonics is also possible but the quantum groups Fourier transform exactly takes us to noncommutative spaces such as $\R^3_\lambda$ which is what we need now.

 Next, we show that the partial derivatives indeed act diagonally on plane waves as
\eqn{deli}{  \del^i=\imath {p^i \over\lambda |\vec p|} \sin({\lambda}|\vec p|)={\imath  \over 2\lambda}{\rm Tr}(\sigma_i\ )}
\eqn{del0}{ \del^0={\imath\mu\over\lambda^2}(\cos({\lambda}|p|)-1)={\imath\mu\over 2\lambda^2}({\rm Tr}-2). }
 The second expressions in each case are just the functions in $\C[SU_2]$ whose evalution on plane waves gives the first expression in each case. Thus  
\[\trace.\psi_{\vec p}=\trace(e^{\imath{\lambda}\vec p\cdot\sigma})\psi_{\vec p}= 2\cos({\lambda}|p|)\psi_{\vec  p}\]
and so forth. It remains to prove the first expressions for the $\del^\mu$.

The full action of the $\del^\mu$ are rather complicated but we need them only for functions of 
\[ X=\imath \vec p\cdot x\]
Then from the relations for the calculus, we find the subcalculus 
\[ (\extd X).X=X\extd X-\imath\nu\theta,\quad \theta X=X\theta-\imath\nu' \extd X\]
where
\[ \nu=\mu p^2,\quad\nu'={\lambda^2\over\mu}. \]
We let
\[ \extd X^n=f_n(X)\extd X+g_n(X)\theta\]
and using the relations above and the Leibniz rule we have
\begin{eqnarray*} \extd X^n&=&(\extd X^{n-1})X+X^{n-1}\extd X=f_{n-1}\extd X.X+X^{n-1}\extd X+g_{n-1}\theta X\\
&=& f_{n-1}X\extd X-\imath\nu f_{n-1}\theta+X^{n-1}\extd X+g_{n-1}X\theta-\imath{\nu'}g_{n-1}\extd X\end{eqnarray*}
hence the recurrence relations
\[ f_n= f_{n-1} X-{\imath\nu'}g_{n-1}+X^{n-1},\quad g_n=g_{n-1}X-\imath\nu f_{n-1};\quad f_1=1, g_1=0\]
These can be easily solved and yield
\[ f_n=-{\imath\over 2\sqrt{\nu\nu'}}\left( (X+\imath\sqrt{\nu\nu'})^n-(X-\imath\sqrt{\nu\nu'})^n\right)\]
\[ g_n=-{\imath\over\nu'} X^n+{\imath\over 2\nu'}\left( (X+\imath\sqrt{\nu\nu'})^n+(X-\imath\sqrt{\nu\nu'})^n\right)\]
or
\[ \extd X^n={\imath\over 2 \lambda |\vec p|}\left( (X-\imath{\lambda}|\vec p|)^n({\mu\over\lambda}|\vec p|\theta+\extd X)+(X+\imath{\lambda}|\vec p|)^n({\mu\over\lambda}|\vec p|\theta-\extd X)\right)  -{\imath \mu\over\lambda^2} X^n\theta.  \]
In particular, we see that
\[e^{-X} \extd e^{X}={\imath\over 2 \lambda |\vec p|} \left(e^{-\imath{\lambda}|\vec p|}({\mu\over\lambda}|\vec p|\theta+\extd X)+e^{\imath{\lambda}|\vec p|}({\mu\over\lambda}|\vec p|\theta-\extd X) \right) -{\imath\mu\over\lambda^2} \theta\]
\[ ={\imath\mu\over\lambda^2}(\cos({\lambda}|\vec p|)-1)\theta+{1\over\lambda |\vec p|}\sin({\lambda}|\vec p|)\extd X\]
which translates into $\del^\mu$ acting as stated on plane waves. Through quantum group Fourier transform this allows one to compute them in principle on any $\psi(x)$. We will not need to do this explicitly, however.

\section{Laplacian and noncommutative Schroedinger's equation}

From the partial derivatives (\ref{deli})-(\ref{del0}) on plane waves, we compute the 3D Laplacians on plane waves:
\[ \nabla^2=\del_i\del^i=-{1\over\lambda^2}\sin^2({\lambda}|\vec p|)\]
which has the correct classical limit $-|\vec p|^2$ as $\lambda\to 0$. Comparing with the expression in momentum space for $\del^0$ we deduce  that
\eqn{d0lap}{ \del^0={\imath\mu\over\lambda^2}\left(\sqrt{1+{\lambda^2}\nabla^2}-1\right)}
on plane waves and hence in general for modes with $\nabla^2\ge  -1/\lambda^2$. Expanding this we find
\eqn{del0sch}{ \del^0\psi =\imath{\mu\over 2 } \nabla^2 \psi+O(\lambda^2)} which is of the form of Schr\"odinger's   equation with respect to an auxiliary `time' variable and a particle with Compton wavelength $\mu$ corresponding to mass $m$,
\eqn{compton}{\mu={1\over m}.}
Our point of view is that $\lambda$ might be of the order of the Planck scale, so if $\mu$ is also of this scale then the effective mass of the particle being described is of the order of the Planck mass.  We are not tied to such a value for either parameter, however.

We now proceed to develop this point of view. Thus let $t$ be a time variable adjoined to the theory, commuting with the position generators and such that
\[ \theta=\extd t\]
which is consistent with our reality assumptions if $t^*=t$. For consistency with the relations in the differential calculus we need
\[0= \extd ([t,x_i])=\theta x_i+t\extd x_i-(\extd x_i)t-x_i\theta \]
so we require 
\[ [t,\extd x_i]=\imath{{\lambda^2\over\mu}}\extd x_i\]
which implies that
\[ (\extd x_i)f(t)=f(t-\imath{{\lambda^2\over\mu}})\extd x_i.\]
In this case for the Jacobi identity  
\[ 0=[\extd x_i,[x_j,t]]+[x_j,[t,\extd x_i]]+[t,[\extd x_i,x_j]]\]
we need
\[ [t,\theta]=\imath{{\lambda^2\over\mu}}\theta\]
Since $\theta=\extd t$ we see that 
\[ f(t-\imath{{\lambda^2\over\mu}})\extd t=(\extd t)f(t) \]
holds as well, which in turn can be used to show that
\[ \extd f(t)= ( \del^t f(t))\extd t;\quad \del^t f(t)\equiv {f(t)-f(t-\imath{{\lambda^2\over\mu}})\over{\imath{\lambda^2\over\mu}}}\]
is necessarily a finite difference operator. Applying $\extd$ again gives the usual anticommutation relations between $\extd t$ and the $\extd x_i$ (as among themselves). In summary, we can adjoin a $t$ variable but for consistency with the spatial noncommuative calculus we will need its calculus to be a noncommutative finite difference one.

The differential on functions $a(x)$ just of $x$ is unchanged:
\[ \extd a(x)=\del^ia(x)\extd x_i+\del^0a(x)\extd t.\]
When we look in the extended algebra we will have functions generated by products of functions $a(x)$ and $f(t)$ and here we find, using the Leibniz rule,
\begin{eqnarray*}
\extd (a(x)f(t))&=& (\del^i a \extd x_i)f+(\del^0 a\extd t)f+a\del^t f\extd t\\
&=&(\del^i a)f(t-\imath{{\lambda^2\over\mu}})\extd x_i + \left((\del^0a)f(t-\imath{{\lambda^2\over\mu}})+a\del^t f\right)\extd t\\
&\equiv & \tilde\del^i(af)\extd x_i+ \tilde\del^0(af)\extd t.\end{eqnarray*}
This defines the partial derivatives $\tilde\del^{\mu}$ acting on general functions in the calculus. Here $\tilde\del^0$ reduces to $\del^0$ as before acting on $a(x)$ and to $\del^t$ acting on $f(t)$ alone. The braided Liebniz rule is evident here on products as the extra shift by $-\imath{\lambda^2}/\mu$ and is typical of noncommutative vector fields. 

The calculus is clearly an unusual one in which the $\nabla^2$ is built into $\tilde\del^0$. Because  we have adjoined the $t$ the  natural formulation of Schroedingers equation  is now
\eqn{extsch}{ \tilde\del^0\psi(x,t)=0}
i.e. functions which are `constant' with respect to the extended differential calculus in the sense that $\extd\psi$ is purely spatial. We can write this more explicitly using the above as
\eqn{finsch0}{{ \imath \mu \over \lambda^2}\left(\psi(x,t)-\psi(x,t-\imath{{\lambda^2\over\mu}})\right)= \del^0\psi(x,t-\imath{{\lambda^2\over\mu}})}
exhibiting $\del^t$ as a finite difference operation in the continuous time variable $t$. Putting
in our previous expression for $\del^0$ this is
 \[\psi(x,t)-\psi(x,t-\imath{\lambda^2\over\mu})= \left(\sqrt{1+\lambda^2\nabla^2}-1\right)\psi(x,t-\imath{\lambda^2\over\mu})\]
 or 
 \eqn{finsch}{  \psi(x,t+\imath{\lambda^2\over\mu})=\left(\sqrt{1+{\lambda^2}\nabla^2}\right)\psi(x,t)}
after a change of variable $t\to t+\imath{\lambda^2\over\mu}$. Writing the left hand side as the action of $e^{\imath{\lambda^2\over\mu}{\del\over\del t}}$ in terms of usual derivatives, and taking $\ln$ of the operators on both sides, we have formally (or not formally on plane waves),
\eqn{logsch}{  {\del\over\del t}\psi=-\imath{\mu  \over 2\lambda^2}\ln\left(1+{\lambda^2}\nabla^2\right)=-{\imath\over 2 m}\left(\nabla^2-{\lambda^2\over 2}\nabla^4+{\lambda^4\over 3}\nabla^6\cdots\right)}
for small $\lambda$. Thus our equation explicitly deforms the usual Schroedingers equation with higher derivative terms. 

Note that we can also expand the left hand side of (\ref{finsch}) in a Taylor series and the right hand side in a binomial series so
 \[ \imath {\del\over\del t}\psi-{\lambda^2\over 2\mu}{\del^2\over\del t^2}\psi+\cdots ={\mu \over 2}\nabla^2\psi -{\lambda^2\mu \over 8}\nabla^4\psi+\cdots.\]
 Now {\em if} we choose $\mu=\lambda$ and keep terms to $O(\lambda^2)$ then this reads in terms of $m$ as
 \[ {\del\over\del t}\psi= -{\imath\over 2 m}\nabla^2\psi-{\imath\over 2m}{\del^2\over\del t^2}\psi+O(({1\over m})^3)\]
which is the 4D Euclidean wave equation for $\phi(x,t)=e^{-\imath {m }t}\psi(x,t)$ in   terms of $\psi$.  Here
\[\left( {\del^2\over\del t^2}+\nabla^2+{m^2}\right)\phi= -2\imath {m}e^{\imath {m}t}\dot \psi+e^{\imath {m}t}\dot{\dot\psi}+e^{\imath {m}t}\nabla^2\psi.\]
On the other hand, $\nabla$ itself is a deformation and at least on plane waves brings its own corrections to the same order so that the above is a formal observation. 

When $\mu=\lambda$ we also have a formal Euclidean aspect in the noncommutative wave operator. In fact we are not proposing an $SO_{1,3}$ or $SO_4$ symmetry in the noncommutative model (in our construction we have preseved spatial rotations only, and induced time from it). However, it is interesting to note that we have the following modified wave operator in the original (spatial) theory, from (\ref{deli})--(\ref{del0}):
\eqn{3Dsquare}{\square^{(3)}=(\del^0)^2+\nabla^2={2\over\lambda^2}(\cos({\lambda}|p|)-1)={1\over\lambda^2}(\trace-2)}
which again has the usual limit $-|\vec p|^2$ as $\lambda\to 0$. The fact that this has such a nice description via the action of $\trace\in \C[SU_2]$  suggests that it is somewhat natural;  if one takes instead  $-(\del^0)^2$ the result is similar but does not have such a simple expresson.   Here $\trace-2$ is the Casimir in $D(U(su_2))=\C[SU_2]\lcross U(su_2)$ as explained in \cite{BatMa:non}.  

\note{This should not necessarily worry us since our goal was to obtain Schroedinger's equation and not a relativistic picture. Nevertheless, working in the 4D picture suggests the Laplacian
\[ \square=(\tilde\del^t)^2+\tilde\del_i\tilde\del^i\]
which reduces to (\ref{3Dsquare}) on functions of $x$ alone.  If one then asks in our noncommutative theory for $e^{\imath{m'c^2\over\hbar}t}\psi(x,t)$ to obey $\square+M^2=0$, say, we have 
in the slowly-varying approximation. This computation at $\lambda\to 0$ is a Euclidean derivation of a  version of Schroedingers equation and works  in the same way as the  more familiar Minkowski space derivation -- it  does not give the diffusion equation. It does however, give a - sign relative to the usual Schroedinger's equation. It is not our strategy in the present paper to recover this but if we wished to, we should that $\theta\to -\theta$ in the conventions above. Also, as noted, we can change $\theta\to \imath\theta$ which is now antihermitian. In this case the theory has a more Minkowski character in  (\ref{3Dsquare}) to obtain the trace, but the 'Schroedinger equation' is now a diffusion equation. These are not directions that we pursue in the present paper. }

\section{Dispersion relations and cut-off}

We can also immediately solve the NCSE on plane waves, using the results of Section~3.  We let $\psi(x,t)=e^{\imath \vec p\cdot x}f(t)$. Then from  (\ref{finsch}) and the value of $\del^0$ on plane waves, we require
 \[ f(t+\imath{\lambda^2\over\mu})=\cos({\lambda}|\vec p|)f(t)\]
 which is solved by
 \eqn{4Dwave}{ \psi_{\vec p}(x,t)=e^{\imath \vec p\cdot x+\imath p^0 t},\quad e^{-{\lambda^2\over\mu}p^0}=\cos({\lambda}|\vec p|),\quad |p|< {\pi\over 2\lambda}.}
Notice that the spatial momentum is bounded above for there to be a primary solution, which is a typical feature of quantum-gravity as discussed in the introduction.  We also have unphysical  solutions ${5\pi\over 2}>|\vec p|>{3\pi\over 2\lambda}$ etc. according to the periodicity of $\cos$, which we consider an artefact of the local coordinate system. Recall that the $p^i$ are local coordinates in momentum space which is actually here a sphere $S^3$ not flat, and this is also the reason for the cut-off in the first place. Note that coordinate system breaks down at $|p|=\pi/\lambda$ so there is still a cut-off effect before that.

Differentiating the $p^0$ equation immediately gives 
\[ e^{-{\lambda^2\over\mu}p^0}{\del p^0\over\del p^i}={\mu\over\lambda}\sin({\lambda}|\vec p|) {p^i\over |\vec p|}\]
and hence  our proposed dispersion relation
\[ |{\del p^0\over\del \vec p}|={\mu\over\lambda}|\tan({\lambda}|\vec p|)|={1\over m\lambda}|\tan({\lambda}|\vec p|)|.\]
i.e. the group velocity is linear for small momentum as expected for a particle of mass $m$ in our nonrelativistic coordinates, but  blows up at the cut-off value.  Note, however, that a detailed analysis of how plane waves in the noncommutative model might be measured experimentally and their group velocity is still needed in order to see if such a naive derivation is justified. In the bicrossproduct model calculation \cite{AmeMa:wav} this was somewhat justified by a 
natural normal-ordering postulate  for the identification of noncommutative expressions with their classical counterparts. In the present model we can in principle do better; we can look at the plane waves in actual representations of the noncommutative algebra on the basis that it is such expectation values that are presumably observed. We propose to use here  the coherent states $|j,\phi,\psi\>$ in which the coordinates behave with minimal uncertainty\cite{BatMa:non} and with expectation values
\[ \<x\>\equiv \<j,\phi,\psi|\vec x|j,\phi,\psi\>=2j\lambda (\sin\phi\, \cos\psi,\sin\phi\, \sin\psi,\cos\phi)\]
for a particle at angle $\phi,\psi$ and radius $2j\lambda$ (which we see is quantised). Explicitly,
\[ |j,\phi,\psi\>=\sum_{k=0}^{2j}2^{-j}\sqrt{\left({2j\atop k}\right)}(1+\cos\phi)^{2j-k\over 2}(1-\cos\phi)^{k\over 2}e^{\imath k \psi}|j,j-k\> \]
in terms of usual spin $j$ states and in the present conventions. In such a state our plane waves have a classical `shadow' as  waves in polar coordinates. For example, 
\[ \<\h,\phi,\psi|e^{\imath \vec p\cdot x}|\h,\phi,\psi\>=\cos\lambda|\vec p|+\imath{\vec p\cdot \<x\> \over\lambda |\vec p|} \sin\lambda|\vec p|\]
which should be compared with the classical value  at this radius of
\[ e^{\imath \lambda(p_1 \sin\phi\, \cos\psi+ p_2\sin\phi\, \sin\psi+p_3\cos\phi)}=e^{\imath \vec p\cdot\<x\>}.\]
The approximation gets better for large spin. This suggests that the noncommutative plane waves would appear as something  with comparable periodicity in the expectation values, and hence comparable group velocity to the extent that the expectation values appear wavelike.  Note also that
\[ |\<j,\phi,\psi | j,\phi',\psi'\>|^2=\left({1\over 2}(1+\cos({\rm angle}(\phi,\psi|\phi',\psi')))\right)^{2j}\]
where the angle is the classical angle between vectors in directions $(\phi,\psi)$ and $(\phi',\psi')$ in polar coordinates. The point of view here is in the spirit of \cite{Pen} but rather than a spin network we use noncommutative geometry. See also \cite{HLS} concerning coherent states. 

\section{Polar coordinates and gauge theory}

The model with time adjoined has in fact a rich noncommutative geometry. Here we note that there is a closed algebra among the $t$ and the Casimir $c=\vec x\cdot\vec x$. These from a commutative subalgebra of rotationally invariant functions varying in time, but as we show now with noncommutative differentials. First of all, we note that
\eqn{dc}{ \extd c=x_i\extd x_i+ (\extd x_i)x_i=2x_i\extd x_i+3\imath\mu \theta}
and $[c,x_i]=0$ implies that
\begin{eqnarray*} [\extd c,x_i]&=&-[c,\extd x_i]=-x_j[x_j,\extd x_i]-[x_j,\extd x_i]x_j\\
&=&\imath\lambda\eps_{ijk}x_j\extd x_k+\imath\mu x_i\theta+\imath\lambda \eps_{ijk}(\extd x_k)x_j+\imath\mu\theta x_i\\
&=&2\imath\lambda\eps_{ijk}x_j\extd x_k-(\imath\lambda)^2\eps_{ijk}\eps_{kmj}\extd x_m+\imath\mu(x_i\theta+\theta x_i)\\
&=&2\imath\lambda\eps_{ijk}x_j\extd x_k+2\lambda^2\extd x_i+\imath\mu(2x_i\theta- \imath{\lambda^2\over\mu}\extd x_i)\\
&=&2\imath\lambda\eps_{ijk}x_j\extd x_k+3\lambda^2\extd x_i+2\imath\mu x_i\theta\end{eqnarray*}
using the relations in the calculus. From this and (\ref{dc})  we find that
\begin{eqnarray}\label{dcc}[\extd c, c]&=&[\extd c,x_i]x_i+x_i[\extd c,x_i]\nonumber\\
&=&-4\lambda^2x_i\extd x_i+3\lambda^2(\extd x_i)x_i+2\imath\mu x_i\theta x_i+x_i3\lambda^2\extd x_i+2\imath\mu c\theta\nonumber \\
&=& 4\lambda^2 x_i\extd x_i+9\lambda^2\imath\mu\theta+4\imath\mu c\theta=2\lambda^2\extd c+4\imath\mu(c+{3\over 4}\lambda^2)\extd t\end{eqnarray}

Meanwhile, from $[t,c]=0$ and $\theta=\extd t$ as generator of $\extd$, we have
\eqn{dct}{ [\extd c,t]=[\extd t,c]={\lambda^2\over\imath\mu}\extd c,\quad [\extd t,t]={\lambda^2\over\imath\mu}\extd t}
to give a closed algebra (\ref{dcc})--(\ref{dct}) among the $t,c,\extd t,\extd c$. As in Section~4 we immediately conclude that
\[ \extd t.f(t)=f(t-\imath{\lambda^2\over\mu})\extd t,\quad \extd c.f(t)=f(t-\imath{\lambda^2\over\mu})\extd c\]
for any function $f(t)$, while the commutation relations with some function $a(c)$ has to be determined by induction. Likewise $\extd f(t)=\del^tf\extd t$ given by a finite difference as in Section~4, while
$\extd a(c)$ has to be determined by induction. 

To this end, let 
\begin{eqnarray*}\extd c^n&\equiv& f_{n}\extd c+ g_n\extd t=\extd c^{n-1}.c+c^{n-1}\extd c\\
&=&
f_{n-1}\extd c.c+g_{n-1}\extd t.c+c^{n-1}\extd c\\
&=&\left( (c+2\lambda^2)f_{n-1}+c^{n-1}-\imath{\lambda^2\over\mu}g_{n-1}\right)\extd c
+ \left(c g_{n-1}+4\imath\mu (c+{3\over 4}\lambda^2)f_{n-1}\right)\extd t\end{eqnarray*}
which gives a recursion relation for $f_n,g_n$ with $f_1=1$ and $g_1=0$. This can be solved to obtain $\extd c^n$, or on any $a(c)$ to compute partial differentials $\del^c$ and $\del^{t}_{|_c}$ defined by
\[ \extd a(c)\equiv (\del^c a)\extd c+(\del^{t}_{|_c} a)\extd t.\]
Here the $|_c$  is to remind us that this is with respect to $c$ and implicit angular coordinates being held constant, which is not quite the same as $\del^0$ in Section~4 where the $x_i$ were being held constant. We find
\[ \del ^c a(c)={a(c+\lambda^2+2\lambda\sqrt{c+\lambda^2})-a(c+\lambda^2-2\lambda\sqrt{c+\lambda^2})\over 4\lambda \sqrt{c+\lambda^2}}\]
\[ \del^{t}_{|_c} a(c)={\mu\over\imath\lambda^2}\left(a(c)+\lambda^2\del^c a(c)-{a(c+\lambda^2+2\lambda\sqrt{c+\lambda^2})+a(c+\lambda^2-2\lambda\sqrt{c+\lambda^2})\over 2}\right)\]
These extend to products by a braided derivation rule and by the same computation as in Section~4 we find
\[ \del^c(af)=(\del^c a)f(t-\imath{\lambda^2\over\mu}),\quad  \del^t_{|_c}(af)=(\del^t_{|_c} a)f(t-\imath{\lambda^2\over\mu})+a\del^t f\]
for $a(c)f(t)$. This gives the partial derivatives and $\extd $ on a general function $\psi(c,t)$ in polar coordinates.  

As an application, we can write our NCSE in polar form as follows. The equation says that $\extd $ is purely in the $\extd x_i$ direction. Writing $\extd a(c)=(\del^ia)\extd x_i+(\del^0a)\extd t$ as in Section~4, we have in view of (\ref{dc}) that
\[ \del^i a(c)=(\del^ca)2x_i,\quad \del^0a(c)=\del^t_{|_c}a+3\imath\mu \del^c a.\]
The latter comes out as 
\begin{eqnarray}\label{d0polar} \del^0a(c)&=&{\imath\mu\over \lambda^2}(\h(1+{\lambda\over\sqrt{c+\lambda^2}})a(c+\lambda^2+2\lambda\sqrt{c+\lambda^2})\nonumber\\
&&\quad\quad +\h(1-{\lambda\over\sqrt{c+\lambda^2}})a(c+\lambda^2-2\lambda\sqrt{c+\lambda^2})-a(c))\end{eqnarray}
using the above results. This compares with (\ref{d0lap}) or (\ref{del0}) computed in our previous plane wave basis. We will not attempt to solve the NCSE here, for one thing one needs to have
suitable proposals for a potential term for, say, a hydrogen atom. Suffice it to say that any such calculation is best done in polar coordinates and (\ref{d0polar})  provides the radial part of the effective Laplacian to be used in (\ref{finsch0}), now on wave-functions $\psi(c,t)$. One may compute for example that
\[ \del^0 ({1\over \sqrt{c+\lambda^2}}) =0\]
where $\rho=\sqrt{c+{\lambda^2}}$ makes for a reasonably nice answer and suggests that this is the appropriate analogue of $r$ in usual polar coordinates. This is the right answer when $\lambda\to 0$ (the radial Laplacian vanishes on $1/r$). Similarly one may check that 
\[ \del^0 c=3\imath\mu,\quad \del^0 c^2=\imath\mu (10c+9\lambda^2),\quad  \del^0\sqrt{c+\lambda^2}=\imath\mu{1\over\sqrt{c+{\lambda^2}}}\]
as expected for $\imath\mu$ times $1/2$ the radial Laplacian  as in (\ref{del0sch}) if $c$ is understood as $r^2$ in the classical limit or more precisely $\rho$ as $r$.

Another application of polar coordinates is for computations in the associated $U(1)$ gauge theory. As mentioned in Section~2 at the basic level a gauge field is a connection $\alpha=\alpha^\mu\extd x_\mu$ with suitable reality properties.  The curvature in the Maxwell theory is just
\[ F_{M}(\alpha)=\extd\alpha=\del^\mu\alpha^\nu\extd x_\mu\wedge \extd x_\nu \]
which may be computed as usual using the partial derivatives. The theory is sensetive to cohomology and a gauge transformation is must the addition of an exact differential. In the noncommutative case we also have the option of a nonlinear $U(1)$ Yang-Mills-type theory with curvature
\[ F_{YM}(\alpha)=\extd\alpha+\alpha\wedge\alpha=\del^\mu\alpha^\nu\extd x_\mu\wedge \extd x_\nu+\alpha^\mu \extd x_\mu\wedge\alpha^\nu\extd x_\nu\]
and which detects homotopy. This needs also the commutation relations between functions and differentials.  We look briefly at  the electrostatic Maxwell case. 

Firstly, we look at gauge fields of the form
\[ \alpha= a(c)\extd t\]
where $a(c)$ does not depend on $t$. This has
\[ F_{M}=\del^ca\extd c\wedge \extd t+ \del^t_{|_c}a(\extd t)^2=\del^ca\, 2x_i\extd x_i\wedge\extd t\]
in view of (\ref{dc}) and $(\extd t)^2=0$ in  the calculus. We recall that the $\extd x_\mu$ anticommute as usual. Thus we have an electric field
\[ E_i=\del^ca(c)2x_i\]
more or less as usual. For example, $a(c)=1/\sqrt{c+\lambda^2}$ gives
\[ E_i=-{x_i\over c \sqrt{c+{\lambda^2}} }\]
which is a inverse square law in the classical limit. Its divergence gives the corresponding charge density. 
 
 More surprisingly, we have also a different way of producing an electric field, namely 
 \[ \alpha=a(c)2x_i \extd x_i=a(c)\extd c- 3\imath\mu a(c)\extd t\]
 in view of (\ref{dc}). We have chosen $\alpha$ here to have purely spatial components in the $\extd x_\mu$ basis but  we have the same conclusion below (with a different coefficient) even if we do not include the second term. The curvature is
 \[ F_{M}=\del^c a(\extd c)^2+\del^t_{|_c}a\extd t\wedge  \extd c-3\imath\mu \del^c a\extd c\wedge\extd t\]
which we compute using
\begin{eqnarray*}
\extd c\wedge\extd c&=&4x_i\extd x_i\wedge x_j\extd x_j+6\imath\mu x_i\extd x_i\wedge\extd t+6\imath\mu \extd t \wedge x_i\extd x_i\\
&=&4x_i\extd x_i\wedge x_j\extd x_j=2[x_i,x_j]\extd x_i\wedge \extd x_j+4x_i[\extd x_i,x_j]\wedge\extd x_j\\
&=&4\imath\lambda\eps_{ijk}x_k\extd x_i\wedge\extd x_j+4\imath\lambda x_i\eps_{ijk}\extd x_k\wedge \extd x_j+4\imath\mu x_i\extd t\wedge\extd x_i\\
&=&2\imath\mu \extd t\wedge\extd c=-2\imath\mu \extd c\wedge\extd t\end{eqnarray*}
using (\ref{dc}) and $(\extd t)^2=0$ in the first line and that $[\extd t,x_i]\propto \extd x_i$ in the second line which produces nothing as $(\extd x_i)^2=0$ for each $i$ (and that $\extd t,\extd x_i$ anticommute as usual). We then use the commutation relations in the calculus and in the algebra. The same observations imply that $\extd c, \extd t$ anticommute. Hence
\[ F_M=-(2\imath\mu\del^c a+\del^0a)\extd c\wedge\extd t=-(5\imath\mu\del^c a+\del^t_{|_c}a)2x_i \extd x_i\wedge\extd t\]
which is again a radial electric field.  Such a potential in the classical case would be absent as $\alpha$ would be pure gauge with zero curvature.  There are clearly many other possibilities to be explored here including time dependent (such as standing wave) solutions. Also in this preliminary analysis we do not discuss source terms for the potentials since this would require a study of the suitable currents produced by matter fields. Finally, all of these remarks are more complicated for the Yang-Mills version.

\appendix
\section{Comparison with bicrossproduct models}

The model above with noncommuting position and commuting $t$ is complementary to the bicrossproduct model where the spacetime $\R^{1,3}_\lambda$ in \cite{MaRue:bic} is
\eqn{kapminksp}{ [t,x_i]=\imath\lambda x_i,\quad [x_i,x_j]=0.}
Here the position coordinates commute and $t$ is noncommutative, but we shall note similar features nevertheless. Some authors write $\kappa=1/\lambda$ as a mass  scale instead. 

This time  the relevant Lie algebra in (\ref{kapminksp}) is the solvable one $b_+$ (say)  and computations are rather easier using normal ordering as explained in \cite{MaRue:bic}. Hence we parametrize the plane waves as
\[ \psi_{\vec p,p^0}=e^{\imath \vec p\cdot  x}e^{\imath
p^0 t},\ 
\psi_{\vec p,p^0}\psi_{\vec p',p^0{}'}=\psi_{\vec p+e^{-\lambda p^0}\vec p',p^0+p^0{}'}\]
which identifies the $p^\mu$ as the coordinates of a nonAbelian group $B_+$  with Lie algebra $b_+$. The group law in these coordinates  is read off as usual from the above product of plane waves. The right-invariant Haar
 measure on $B_+$ in these coordinates is the usual $\extd^4 p$ so the quantum group Fourier transform\cite{Ma:ista} reduces to the usual one but
 normal-ordered,
 \[ \CF(f)=\int_{\R^4}\extd^4 p\ f(p) e^{\imath \vec p\cdot x}e^{\imath p^0 t}.\]
 As before, the action of elements of $\C[B_+]$ define
differential operators on $\R^{1,3}_\lambda$ and these act diagonally on plane waves.    
 
There is also known to be a natural differential calculus with
\[ (\extd x_j)x_\mu=x_\mu\extd x_j,\quad (\extd
t)x_\mu-x_\mu \extd t=\imath\lambda \extd x_\mu\]
which we see already has a generator with $\theta=\extd t$; so we do not need to adjoin a further $t$ in this model. The unitarity or  $*$-structure is $x_i^*=x_i$, $t^*=t$ and the same for their differentials. The calculus recovers  the partial derivatives \cite{MaRue:bic}
\[ 
\del^i\psi=:{\del\over\del x_i}\psi(x,t):=\imath p^i.\psi\]
\[ 
\del^0\psi=:{\psi(x,t+\imath\lambda)-\psi(x,t)\over\imath \lambda}:={\imath\over\lambda}(1-e^{-\lambda p^0}).\psi\]
for normal ordered polynomial functions $\psi$ or as shown in terms of the action of the momenta $p^\mu$.  It was shown in \cite{AmeMa:wav} that by using adjusted derivatives $L^{-\h}\del^\mu$ where
\[ L\psi=:\psi(x,t+\imath\lambda):=e^{-\lambda p^0}.\psi\]
 the  4-D Laplacian $\square=L^{-1}((\del^0)^2-\sum_i (\del^i)^2)$  acts on plane waves as 
\eqn{kappacas}{\square=-{2\over\lambda^2}(\cosh(\lambda p^0)-1)+{\vec p}^2e^{\lambda p^0}=e^{\lambda p^0}\left(\vec p^2-{1\over\lambda^2}(1-e^{-\lambda p^0})^2\right)} 
 which is the action of the Casimir of the bicrossproduct quantum group. The first expression should be compared with (\ref{3Dsquare}) in our model above, as a hyperbolic version of it. 
  
We recall that this bicrossproduct Poincar\'e quantum group   $U(so_{1,3})\rlbicross \C[B_+]$  in terms of translation generators $p^\mu$, rotations $M_i$ and boosts $N_i$, is \cite{MaRue:bic}
\[   [p^\mu,p^\nu]=0,\quad [M_i,M_j]=\imath\eps_{ij}{}^kM_k,\]\[ 
[N_i,N_j]=-\imath\eps_{ij}{}^kM_k,\quad [M_i,N_j]=\imath\eps_{ij}{}^kN_k\]
 \[ [p^0,M_i]=0,\quad
[p^i,M_j]=\imath\eps^i{}_{j}{}_kp^k,\quad [p^0,N_i]=-\imath p_i,\]
as usual, and the modified relations and coproduct
\[ {}[p^i,N_j]=-{\imath\over 2}\delta^i_j \left({1-e^{-{2\lambda p^0}}\over
\lambda}+\lambda{\vec p}^2\right)+\imath\lambda p^ip_j,\]
\[  \Delta N_i=N_i\tens 1+e^{-{\lambda p^0}}\tens
N_i+\lambda\eps_{ijk}p^j\tens M_k,\]
\[ \Delta p^i=p^i\tens
1+e^{-{\lambda p^0}}\tens p^i\]
 and the usual linear ones on  $p^0,M_i$. We raise and lower $i,j,k$ indices using the Euclidean metric. It follows from the general theory of bicrossproducts that this Hopf algebra acts on $U(b_+)=\R_\lambda^{1,3}$.

 \begin{figure}
 \[ \kern-30pt  \includegraphics{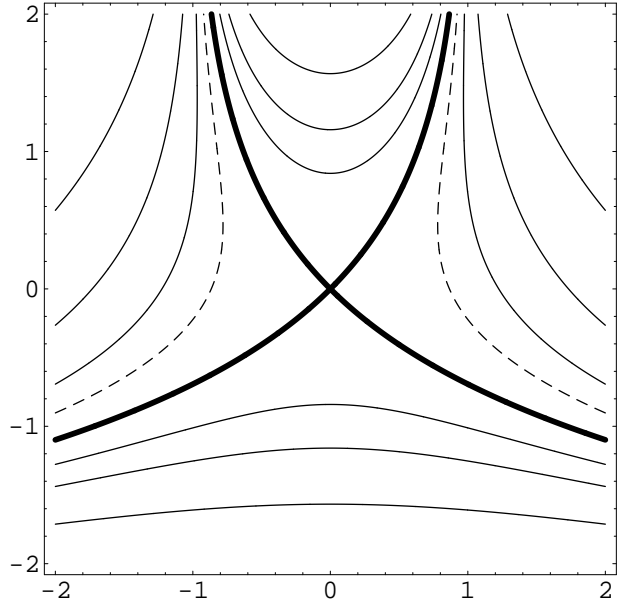}\]
 \caption{Deformed mass-shell orbits in the  bicrossproduct curved momentum space for $\lambda=1$}
 \end{figure}

 Part of the motivation for this  model  \cite{MaRue:bic} consisting of noncommutative spacetime and associated bicrossproduct quantum group acting covariantly on it  was a  previous  `$\kappa$-Poincar\'e'  version of the Hopf algebra alone   obtained \cite{LNRT:def}  in another context (by contraction of  $U_q(so_{2,3})$). The bicrossproduct model should not be confused with this, however, because its generators and relations are fundamentally different and have very different physical content; for example the Lorentz generators  in \cite{LNRT:def} do not close among themselves but mix with momentum.  Moreover, prior to \cite{MaRue:bic} there was either no action of $\kappa$-Poincar\'e on any spacetime  or it was taken to act on classical Minkowski space with inconsistent results (there is no such covariant action).

 Also key and fundamentally different in the bicrossproduct model  from `$\kappa$-Poincar\'e' is that in the bicrossproduct case  the  Lorentz group is actually undeformed; rather  the deformation is in a nonlinear but entirely geometrical action of it on the `curved' momentum group $B_+$. This is as part of the solution of a non-linear set of `matched pair' equations \cite{Ma:pla}   (the other part of the matched pair is a `back reaction' of $B_+$ on the manifold of the Lorentz group).  Because of  this fundamental  difference it would be a mistake to view the bicrossproduct quantum group as merely a `change of basis' from $\kappa$-Poincar\'e. In particular,  because of the classical geometry behind it one can see what is going on in terms of the curved momentum space, as shown in Figure~1, which is a contour plot of $p^0$ against $|\vec p|$. The  nonlinear action of the Lorentz group means that   Lorentz group orbits in $B_+$ are now deformed hyperboloids. Because neither group here is compact one expects again from the general theory of bicrossproducts to have limiting accumulation regions and we see that indeed the $p^0>0$ mass shells are now cups  with almost vertical walls, compressed into the vertical tube $ |\vec p|< \lambda^{-1}$. In other words, the 3-momentum is bounded above by the Planck momentum scale (if $\lambda$ is the Planck time), but this does not appear as a new `second bound' in addition to Einstein's postulate on the speed of light as sometimes claimed in the literature, but rather a deformation of it.

Such accumulation regions  were already  visible  in the simplest `Planck-scale Hopf algebra' \cite{Ma:pla} from the 1980s under a different point of view. Under this point of view $p^i$ above are position and not momentum coordinates and the quantum group is the algebra of obervables for a quantum particle moving on oribits under (in the present case Lorentz) group action. The flows for this action are geodesics on the orbits, which fit together to a natural 4D space that could be called a `pseudo-black hole'. Here the physical region of Figure~1 is the orbits that come in from spatial infinity and remain outside the tapered cylinder of radius $1/\lambda$.  One type of orbit comes in at large position, bends upwards and asymptotically approaches the cylinder from the outside for large $t$ (which points upwards), much as for an event horizon. The other type (dashed) crosses the asymptote and approaches it from the inside. The detailed geometry of this setup will be presented elsewhere. Such coordinate singularities are a generic feature of the nonlinear  `matched pair' equations behind the model and is the reason that they were proposed in \cite{Ma:pla} as a toy version of Einstein's equations.

\begin{figure}
 \[ \kern-30pt  \includegraphics{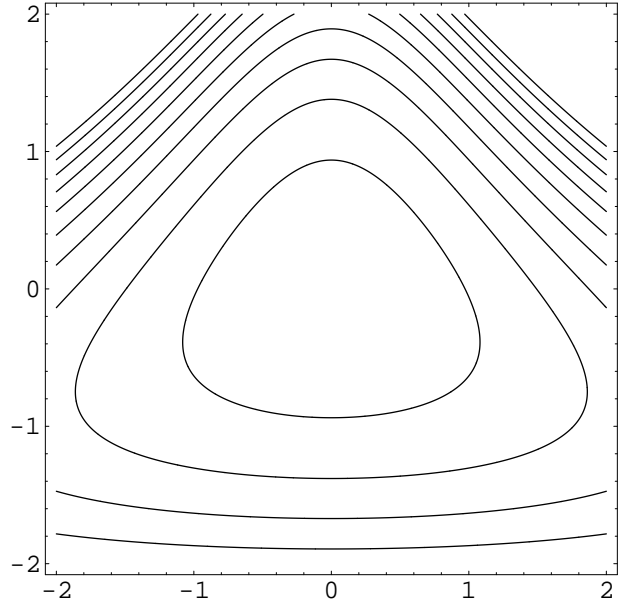}\]
 \caption{Deformed spherical orbits in the  3D bicrossproduct model for $\lambda=1$}
 \end{figure}

Finally, we point out what does not appear to be well-known that the above bicrossproduct is part of a family of which the  3D Euclidean  bicrossproduct $\C[B_+]\lrbicross U(su_2)$ was already obtained in the 1980s in \cite{Ma:the,Ma:mat,Ma:hop} actually as a Hopf-von Neuman algebra and which has the following algebraic structure. Firstly, $B_+$ is now a 3D version of the same solvable group, with Lie algebra 
\eqn{3dkappa}{ [x_3,x_i]=\imath\lambda x_i,\quad [x_i,x_j]=0}
for $i=1,2$. This Lie algebra (with generators denoted $\{l_i\}$) and the required nonlinear solution of the matched pair equations are  in \cite{Ma:mat}. The original interpretation of $\C[B_+]\lrbicross U(su_2)$ was different (namely  particles moving in orbits in $B_+$ as position space) but there is of course nothing stopping 
one considering it as a deformation of the group of motions on $\R^3$. The only difference is to denote the generators of $\C[B_+]$ by the symbols $p^i$, which we also combine with a cosmetic change to a logarithmic coordinate and explication of the deformation parameter, i.e.
\[ e^{-\lambda p_3} ={1\over X_3+1},\quad \lambda p_i={X_i\over X_3+1} \]
in terms of the $B_+$ coordinates $\{X_i\}$ written in lower case in \cite{Ma:ista,Ma:book}. We reserve $x_i$ instead for the auxiliary noncommutative space (\ref{3dkappa}) on which the quantum group necessarily acts. Then the bicrossproduct   has the form 
\[ [ p_i, p_j]=0,\quad   [M_i,M_j]=\imath\eps_{ij}{}^kM_k\]
\[ [M_3,p_j]=\imath\eps_{3j}{}^kp_k,\quad [M_i,p_3]=\imath\eps_{i3}{}^kp_k\]
as usual, for $i,j=1,2,3$, and the modified relations  
\[ [M_i,p_j]={\imath\over 2}\eps_{ij}{}^3\left({1-e^{-2\lambda p_3}\over \lambda}-\lambda\vec{p}^2\right )+\imath\lambda \eps_{i}{}^{k3}p_jp_k\]
for $i,j=1,2$ and $\vec p^2=p_1^2+p_2^2$. The coproducts are 
\[ \Delta M_i=M_i\tens e^{-\lambda p_3}+\lambda M_3\tens p_i+1\tens M_i\]
\[ \Delta  p_i= p_i\tens e^{-\lambda p_3}+1\tens p_i\]
for $i=1,2$ and the usual linear ones for $p_3,M_3$. 
 
The deformed spherical orbits under the nonlinear rotation in $B_+$ are constant values of the Casimir for the above algebra. This is was found in \cite{Ma:mat} as
\[ {2\over\lambda^2}(\cosh(\lambda p_3)-1)+\vec p^2 e^{\lambda p_3}\]
in our present coordinates,  which is the Euclidean precursor to (\ref{kappacas}) above. These deformed orbits are shown in Figure~2. The model here is a Euclidean inhomogeneous one.  The noncommutative differential geometry on (\ref{3dkappa}) is broadly similar to the 4D case.

\baselineskip 14pt
\bibliographystyle{unsrt}
\bibliography{biblio}

\end{document}